\documentclass[12pt]{article}
\usepackage{epsfig}
\newcommand{\fm}{\mbox{$.\!^m$}}
\newcommand{\fd}{\mbox{$.\!^d$}}
\newcommand{\fh}{\mbox{$.\!^h$}}

\begin{document}
\begin{center}
{\Large \bf Idling Magnetic White Dwarf\\[2mm]
 in the Synchronizing Polar BY Cam.\\[2mm]
 The "Noah-2" Project}\\[2mm]
\rm
{Ivan L. Andronov$^{1,2},$
\footnote{tt\_ari@ukr.net; il-a@mail.ru; il-a@mail.od.ua; uavso@pochta.ru; andronov@osmu.odessa.ua},
Kirill A. Antoniuk$^3,$
Vitalii V. Breus$^2,$\\
Lidia L. Chinarova$^2,$
Won Young Han$^4,$
Young Beom Jeon$^4,$\\
Yonggi Kim$^{5,6},$
Sergey V. Kolesnikov$^2,$
Joon Young Oh$^5,$\\
Elena P. Pavlenko$^3$ and
Nikolay M. Shakhovskoy$^3$\\[2mm]}
\end{center}
$^1$ {Odessa National Maritime University, Mechnikov str., 34, Odessa 65029, Ukraine}\\
$^2$ {Astronomical Observatory and Department of Astronomy, Odessa National University,\\
             T.G.Shevchenko Park, Odessa 65014, Ukraine}\\
$^3$ {Crimean Astrophysical Observatory, Nauchny 98409, Ukraine}\\
$^4$ {Korea Astronomy Observatory and Space Science Institute, Daejeon 305-348, Korea}\\
$^5$ {University Observatory, Chungbuk National University, Cheongju 361-763, Korea}\\
$^6$ {Institute for Basic Science Research, Chungbuk National University, Cheongju 361-63, Korea}\\
\begin{abstract}
Results of a multi-color study of the variability of the magnetic
cataclysmic variable BY Cam are presented. The observations were obtained at the Korean 1.8m and Ukrainian 2.6m, 1.2m and 38-cm telescopes in 2003-2005, 56 observational runs cover 189 hours. The variations of the mean brightness in different colors are correlated with a slope dR/dV=1.29(4), where the number in brackets denotes the error estimates in the last digits. For individual runs, this slope is much smaller ranging from 0.98(3) to 1.24(3), with a mean value of 1.11(1). Near the maximum, the slope becomes smaller for some nights, indicating more "blue" spectral energy distribution, whereas the night-to-night variability has an "infrared" character. For the simultaneous UBVRI photometry, the slopes increase with wavelength from dU/dR=0.23(1) to dI/dR=1.18(1). Such wavelength dependence is opposite to that observed in non-magnetic cataclysmic variables, in an agreement to the model of  cyclotron emission. The principal component analysis shows two (with a third at the limit of detection) components of variablitity with different spectral energy distribution, which possibly correspond to different regions of emission. The scalegram analysis shows a highest peak corresponding to the 200-min spin variability, its quarter and to the 30-min and 8-min QPOs. The amplitudes of all these components are dependent on wavelength and luminosity state. The light curves were fitted by a statistically optimal trigonometrical polynomial (up to 4-th order) to take into account a 4-hump structure. The dependences of these parameters on the phase of the beat period and on mean brightness are discussed. The amplitude of spin variations increases with an increasing wavelength and with decreasing brightness.
The linear ephemeris based on 46 mean minima for 2003--2005  is HJD $2453213.010(3)+0.137123(3)E.$
 The extensive tables of the original observations and of results of analysis are published in an electronic form.
The nearby star GSC 4081-1562 was found to be an eclipsing red variable.
\end{abstract}
{\it Keywords:
Cataclysmic variables, Close binaries, X-Ray binaries, Individual stars (BY Cam)}
\newpage

\section{Introduction}
Cataclysmic variables are close binaries with a red dwarf filling its Roche lobe and a white dwarf. Accretion structure is strongly dependent on the characteristics of the flow as well as on the magnetic field of the compact star (see e.g. monographs by Warner [55] and Hellier [20]). 

The object H0538+608 = BY Cam was discovered as an X-ray source by Forman et al. [18].
Remillard et al. [44] classified it as an AM Herculis-type system. Mason et al. [34, 35] provided near simultaneous polarimetry and phase-resolved spectroscopy and noticed extreme variations in the accretion geometry. Andronov and Fuhrmann [11] suggested that these variations are due to a small asynchronism between the spin and orbital rotation of the magnetic white dwarfs. So there are slow periodic changes of the magnetic axis of the white dwarf (and corresponding changes of the accretion structure) in respect to the red dwarf with a period of 14\fd52 [50]. These changes are sometimes called "idlings" (or, alternately, "scrollings", "slewings"). In classical polars, the white dwarf is synchronized with the orbital motion, but the orientation of the magnetic axis may undergo cyclic changes in respect to the line of centers of the rotating cataclysmic binary system. Such changes were called "swingings" (Andronov [2]), contrary to "idlings" in the asynchronous polars caused by different orbital and spin periods of the white dwarf.
With changing orientation of the magnetic axis in respect to the secondary (donor) star, also changes the location and other characteristics of accretion columns above the photosphere of the white dwarf.

From the cyclotron humps in the infrared spectra, Cropper et al. [17] estimated magnetic field strength of the white dwarf of 41MGs, which is larger than in AM Her itself.

 This puts the object between relatively numerous groups of DQ Her -- type stars (or intermediate polars) with rapidly rotating white dwarf and the AM Her -- type stars (or classical polars) which are in synchronism [39]. Another asynchronous system was Nova Cyg 1975 = V1500 Cyg [49], thus the desynchronization may be caused by a Nova outburst. Two more recent asynchronous polars are known (V1432 Aql and CI Ind). Sometimes these stars are called as the BY Cam -- type stars (e.g. Mason [33], Honeycutt and Kafka [23]).

According to the last "live" electronic version of the catalogue (dated 01.09.2006) by Ritter and Kolb [46], the number of classical, asynchronous and intermediate polars is 77, 4 and 38 (from a total number of cataclysmic variables of 641), respectively. So asynchronous polars are relatively rare objects. To study BY Cam, Silber et al. [50] have organized the "Noah Project" with a large flow of data obtained during more than 40 nights. Mason et al. [36] have succeeded to explain contradictions between different periods using the model of alternately changing active poles of the white dwarf. In this paper, we report results from the "Noah-2" project initiated to study different types of variability of this interesting object. During the "Noah-2" project, we have obtained 56 nightly runs of observations, more than 40 nights initially planned for the "Noah-1" project, and more attention is paid to the color information. The extensive tables of individual observations and the model parameters are presented electronically.

\section{Observations and Comparison Stars}

\subsection{Observations}
  The observations were made in few observatories:
\begin{itemize}
\item Bohyunsan Optical Astronomy Observatory (BOAO) of the Korea Astronomy Observatory and Space Science Institute (KASI). The observations have been obtained with a thinned SITe 2k CCD camera attached to the 1.8m telescope. The instrumental V and R systems have been used in the mode of alternatively changed filters. To determine instrumental magnitudes of stars, the IRAF/DAOPHOT package [37] has been used.
\item Crimean Astrophysical Observatory (CrAO) with the 2.6m reflector ZTSh with the "wide R" filter (partially overlapping with V) equipped with a new version of the photometer-polarimeter, which allows to determine both linear and circular polarization. The computer program for the data reduction was introduced by Breus et al. [16]. 
\item CrAO with the 1.25 m reflector AZT-11 UBVRI photometer-polarimeter [43].
\item CrAO with the 38cm reflector K-380. Some results of this monitoring had been published separately [40]. However, in the present study, we have re-analyzed these data with new ones to make a more complete study.
\end{itemize}

The VR brightness of the comparison stars based on the BOAO observations is shown in Table 1 (see a more detailed discussion below).

The journal of observations is presented in Table 2 (available only electronically). In an addition to usually presented values, we have added the phase for the mid-run according to the "beat period" ephemeris [40]
\begin{equation}
T_{phot.beat min.}(HJD) = 2 453 089.2473 + 14.568\cdot E 
\end{equation}
where the initial epoch corresponds to the minumum of mean brightness. The light curves are modulated with this period, thus the corresponding phase is important to compare light curves from different beat cycles (see below).

Altogether 56 nights of observations were obtained, 7 from which are with changing filters (VR, BOAO, 1.8m, $33^{\rm h}$), 1 in V (1.8m), 4 with 3-sec photometry (R, ZTSh, 2.6m), and 1 in a mode of simultaneous 5-color 10-sec photometry (UBVRI, AZT-11, 1.25m, 4\fh7). The rest 47 runs are of monochromatic CCD (R, K-380, 0.38m) photometry. 

\subsection{Comparison stars}

The photometric BV standards in the field have been published by Henden and Honeycutt [22], from which some were chosen as comparison star candidates by Sumner [52]. They are marked in Fig.~1 by "H" and "S" with a number, respectively. Henden [21] has published another (more extended) sequence also in BV only, with magnitudes, which are slightly different from that of Henden and Honeycutt [22]. For the R photometry, no published standards have been found.

The brightness of the comparison star C0 (GSC 4094:2109) was determined using the UBVRI photometer-polarimeter by linking to the standard star BD$+27^\circ 2120$ [38]: U=10\fm20, B=10\fm11, V=9\fm95,R=9\fm74 I=9\fm67 [7].
The magnitudes of the comparison star C7 (GSC 4081:280) from our measurements are U=15\fm405, B=13\fm468, V=11\fm762, R=10\fm390, I=9\fm359. The BV magnitudes are in a fairly good agreement with another estimate B= 13\fm473, V=11\fm761 [22]. This star was used for photoelectric photometry at the 1.25m telescope in CrAO. For the CCD photometry obtained at the 1.8m telescope of BOAO, we have used the stars C1-C5 for the "artificial" comparison star [8, 26]. At the 38cm telescope of CrAO, the star C11 was used as the comparison star, and C8 as the check star. To determine relative brightness of comparison stars used at different telescopes, we have measured stars C7-C20 on 10 pairs of CCD images obtained in the BOAO. As both BOAO V and R instrumental systems are close to the standard ones [27], no color correction was applied.

The adopted brightness of the comparison star C5 (GSC 4081:509) is V=13\fm373 and R=12\fm595. Other published values are V=13\fm408$\pm$0\fm008 [22] and V=13\fm392$\pm$0\fm032 [21] are in a reasonable agreement with our data.
This star was set as the "main" one, as it has the color index closest to that of the variable star. The magnitudes of C11 (GSC 4094:467) determined by Pavlenko [40] are V=14\fm55, R=13\fm96. These values have been used for the K-380 and ZTSh data. This V magnitude coincides with the USNO value  and is intermediate between our BOAO-based estimate V=14\fm514 and the value 14\fm572 [22] and coincides with the USNO value. However, the magnitude in R differs from the BOAO value more significantly: R=13\fm68. Because of the difference between the photometric systems used, there still may be some systematic shifts, which are much smaller than the $\sim1^{\rm m}$ amplitude. Thus we analyse the runs obtained using different telescopes separately.

All original observations (HJD, magnitude) are presented electronically in  Tables 3, 4 (V, R data obtained in the BOAO), 5 (ZTSh R data), 6 (UBVRI brightness, color and principal components, see section 6) and 7 (K-380 R data).

The r.m.s. accuracy estimates for a single BOAO measurement from the "brightness-scatter" dependence are 0\fm004-0\fm007 (V) and 0\fm003-0\fm013 (R) for the maximum and minimum brightness, respectively. For the K380 data, the accuracy is 0\fm01-0\fm03 [40]. For 3-sec integrations at the 2.6m telescope, the accuracy is 0\fm02-0\fm03.

\subsection{New variable star GSC 4081:562}
The star C4 (GSC 4081:562, $\alpha_{2000}=5^{\rm h}42^{\rm m}41^{\rm s},$
$\delta_{2000}=60^\circ52'19")$ was suspected to be an eclipsing variable in a range 13\fm27-13\fm45 in V and 11\fm43-11\fm56 in R (based on nightly mean values from the BOAO data). An amplitude is small $\Delta V=0\fm18,$ $\Delta R=0\fm13,$ the color index varies from V-R=1\fm84 at the maximum to 1\fm89 at the minimum and an eclipse duration is of few days. Despite the formal error estimates are smaller than 0\fm003, in the final publication, we round them to 0\fm01. The minimum occurred at JD 2453054. A very large color index B-V=1\fm856 argues for a late--type red component, possibly a giant. Using the magnitudes at the maximum and minimum, we have determined the magnitudes of the eclipsed body: V=15\fm32 and R=13\fm82. The resulting color index V-R=1\fm50 corresponds to a hotter star. During the night, some photometric waves were observed with a typical peak-to-peak amplitude $\sim0\fm02,$ which may be caused by difference in extinction for stars with so different color indexes. A detailed study of this object will be published separately.

\section{Active vs inactive states}

The prototype polar AM Her is characterized by transitions between high and low luminosity states, which Hudec and Meinunger [24] called as an "active" and "inactive" ones, respectively. The star was mostly in the high state during the period 1928-1976, they have found 7 "excursions" to the low state of a duration of 130-700$^{\rm d}.$
Recently Kafka and Honeycutt [25] have detected 10 inactive states in AM Her during 1990-2004, which seem to be more abundant than in previous decades. Based on their studies of magnetic and non- magnetic variables, they interpret relatively rare low states as resulted by different mechanisms for the systems above and below the period gap. For long-period systems, they are resulted by starspots at the secondary at the inner Lagrangian point [31], whereas for short-period systems, by the irradiation feedback on the secondary star [14, 29].

Another mechanism which is effective in a case of channelizing of the accretion flow close to the inner Lagrangian point, is the "magnetic valve" [1] combined with a "swinging dipole" [2], i.e. changing orientation of the magnetic axis of the white dwarf at a time-scale 3-11 yr. This is definitely not the case of BY Cam, where the spin-orbital synchronization has been not yet reached. However, in this case one can suggest variations of the latitude of the magnetic axis (and, correspondingly, of the axis of the accretion column). Similar "precession-like" variations may be expected in a case of non-coincidence of the spin and orbital rotational axes of the white dwarf [6]. Accretion variations at a few year-scale are seen in long-term brightness variations [15] and seasonal variations of the outburst cycle length of dwarf novae [13] and are usually interpreted as a solar-type magnetic cycles [45]. Such changes may also be partialy caused by minor variations of the orbital separation due to a third body which cause a 'trigger-like' amplified response in the accretion rate [10]. All these mechanisms produce quasi-perodic variations of the luminosity at a few year (or decade) time-scale and may act simultaneously.

Turning now to BY Cam, the analysis of the published AFOEV (2005) data shows a wave with relatively sharp minima at JD~2451800 and 2453500, with a difference of $\sim1700^{\rm d}$ and a full amplitude 0\fm8. Similar conclusions may be done from the AAVSO data, which are slightly shifted in time.

Honeycutt and Kafka [25] suggested a 145$^d$ periodicity from the CCD data obtained using the 41-cm Robotic Telescope (RoboScope) from 1991 to 2004. Unfortunately, these sources of data are not corrected to 198-min variability with a large $(\sim 1^{\rm m})$ amplitude, even in a simplest form of a mean phase curve. Thus the scatter of the light curve is highly affected by this kind of variability of the star.

The complete light curve from our observations in the filters V and R is shown in Fig.2. One may note that the abrupt $\sim 1^{\rm m}$ variability of the nightly mean brightness sometimes occurs at a time scale of $\sim 1^{\rm d}$ (from our data), whereas "slow" transitions from "bright" to "faint" luminosity states and {\em vice versa} last $\sim16^{\rm d}$ and $\sim35^{\rm d},$ respectively. Szkody et al. [53] detected a system at the "low" luminosity state $\sim17\fm5.$ Thus we may conclude that all our measurements correspond to a "high" state (according to usual terminology), and thus we will use terms "bright" and "faint" to distinguish between the runs with a mean magnitude difference of $\sim1^{\rm m}.$

\section{Brightness-Color Dependence}
\subsection{Nightly mean values}
The Korean CCD data have been mainly obtained in a two-color (alternate switching filters VR) for an analysis of the color variations. To determine "quasi-simultaneous" VR data for the calculation of V-R (similarly to BG CMi [27] and MU Cam=1RXS J062518.2+733433 [28]), we have used a local cubic fit for interpolation of the brightness in V for the moments of observations in another filter R (and {\em vice versa}), see Andronov and Baklanov [8] for details. Thus the number of "quasi-simultaneous" data is slightly less than twice the minimum number of points (from two filters), as it is well illustrated by Table 2 (Journal of observations). One may note that generally the mean value of $\langle V-R\rangle\neq\langle V\rangle-\langle R\rangle,$ if different sets of points are used (i.e. for $\langle R\rangle$ we used original values of the brightness, whereas for $\langle V-R\rangle$ only the interpolated ones may be used).

From 7 nights, two correspond to a "faint luminosity" state, and five - to a "bright luminosity" state. Fig. 3 shows a dependence for the nightly mean values. It has to be noted, that the error estimates for the color index  $\langle V-R\rangle$ is much smaller then of the  $\langle R\rangle,$ because the variations in two filters V and R are highly correlated, thus a color index has a much smaller amplitude (different for different nights) then a brightness in an individual filter. The high correlation $(\rho=0.97\pm0.11)$ is present, with a best fit dependence:
\begin{equation}
\langle V-R\rangle = 0\fm939(30) - 0.219(24)\cdot(\langle R\rangle-13.98) 
\end{equation}
Here we have used the form $y(x)=\bar{y}+b\cdot(x-\bar{x})$ of a linear depencence, which minimizes the error estimate of the zero point [56]. The amplitude of the variations of the nightly mean values (which mainly characterize the luminosity changes) in R is larger then in V, with value of the slope of $dV/dR=\Delta R/\Delta V=0.88(2)$ (or $dR/dV= 1.29(4)$). Sometimes in the literature, such a "slope" is called as a "gradient".

For small-amplitude variability seen e.g. in the intermediate polars and nova-like variables, the slope dV/dR may be converted to the color excess $\Delta(V-R)=-2.5\lg({\rm dV/dR})$ in respect to the mean color index [54].

However, in BY Cam, the amplitude is large, thus we prefer to use slopes instead of color excesses for V-R and other colors.

\subsection{Nightly run values}

The slope value 1.29(4) characterizes only the mean brightness level. For faster variability, we have determined the slopes for the individual nights. 
They are listed in Table 8, which is available only electronically.
They vary from 0.98(3) to 1.24(3), with a value 1.200(8) for all 7 nights. The latest value is dominated by the variations of the mean brightness. To eliminate this strong dependence described above, we have removed corresponding nightly mean values, thus obtaining a significantly smaller slope of 1.105(11), which is consistent with mean slopes both for combined "all bright" and "all faint" runs.


One may also suggest a minor effect of changing slope during the night, in a sense that $\Delta R/\Delta V$ is smaller for the bright parts of the $3^{\rm h}$ light curve, and is larger near the minimum. An examination of the $"V-\bar{V}, R-\bar{R}"$ diagram shows that this change of slope occurs at $V-\bar{V}\approx0\fm10.$ For corresponding branches of the diagram and all data, we have obtained the values of slope of 0.85(4) and 1.23(2).
The latter value is much more closer to the one 1.29(4) for the mean brightness, indicating that the main source of emission (differently seen at different mean values of brightness) has much lower color temperature, than the one responsible for the $3^{\rm h}$ brightness variability. At the maximum, the color temperature becomes even larger.

This may be illustrated also using the 4-harmonic fits to our light curves (Fig. 4). The "V,R" diagram shows a relatively narrow zone. However, sometimes there are "loops", as the variations in V and R are not precisely simultaneous. 

The "infrared" dependence of amplitudes of the spin variations on wavelength is opposite to a "blue" one observed in non-magnetic cataclysmic variables, e.g. TT Ari [54], where a $\sim3^h$ variability is interpreted by rotation of the accretion disk. The present result for BY Cam is in a qualitative agreement to the model of cyclotron emission with a maximum in the infrared: with an increasing mass transfer, increases the cyclotron emission, so the system becomes more "red". The spin variability is due to periodic changes of the orientation of the accretion column in respect to the line of sight. The flux is dependent on additional parameter - the angle between the column and the line of sight, also there may be eclises of the bottom part of the column. A quantitative modeling of the observed dependencies needs additional polarimetric and spectral observations, as the number of model parameters (mass of the white dwarf, accretion rate, orientation and thickness of accretion columns, density distribution, strength and configuration of the magnetic field, orbital inclination) exceeds the number of dependencies. This may be a subject of another paper.

\section{Spin variability}

\subsection{Periods and their origin}

As the emission lines undergo significant changes with phases of both spin and beat periods, they are not good indicators of the orbital motion. A linear regression to the timings of inferior conjunction of the red dwarf gives 
\begin{equation}
T_{SP}(HJD) = 2 451 137.3952(36) + 0.13975284(14)\cdot E 
\end{equation}
[48], known as the "201-minute" (or "orbital") period. This differs from previously published estimates. The spectroscopic period is distinctly different from the photometric one with an ephemeris for 1991-2004:
\begin{equation}
T_{phot.min.}(HJD) = 2 448 486.98(3) + 0.137120(2)\cdot E 
\end{equation}
[23]. This difference was explained by Mason et al. [36] as a $"2\omega-\Omega"$ frequency, where $\omega$ is a spin frequency of the white dwarf, and $\Omega$ is an orbital frequency, following the model by Wynn and King [57] for intermediate polars.
This photometric period coincides with that from the more recent ephemeris
\begin{equation}
T_{phot.min.}(HJD) = 2 453 022.268 + 0.137120\cdot E 
\end{equation}
(Pavlenko 2006). Using all available timings for the main maximum, Pavlenko et al. [41] have published an ephemeris
\begin{equation}
T_{spin}(HJD) = 2 453 089.307 + 0.1384274\cdot E. 
\end{equation}
 We also will use the "beat" ephemeris correponding to a synodic (spin - orbital beat) period $P_{beat}=(P_{spin}^{-1}-P_{orb}^{-1})^{-1}$
(Eq.1). These two ephemerids have been used for further study.

\subsection{4-harmonic model}

For periodic light curves, usually a multi-harmonic fit
\begin{equation}
m(t) = C_1 + \sum_{j=1}^s (C_{2j}\sin(2\pi jt/P)+C_{2j+1}\cos(2\pi jt/P)). 
\end{equation}
is applied with a determination of statistically optimal value of the degree of the trigonometric polynomial $s.$  Here $m(t)$ is a signal at the moment $t,$ $P$ is an adopted period (we have used $P=0\fd137120$ published by Pavlenko [40]), and $C_k,$ $k=1..(2s+1)$ are parameters determined using the least squares method.
A description of working formulae for the computer program was presented by Andronov [3].
Suggesting a 4-hump model of the light curve [40], which is often seen in our data, we have 
applied the fixed value of $s=4.$ 

The individual light curves for different telescopes are shown in Fig. 5-7 together with the corresponding fits. The characteristics of the maxima and minima will be discussed in section 7. They are presented in the electronic Table 9.

In Fig. 5, the fits are shown for the V, R and V-R curves obtained in BOAO. They show significant variability of the shape, which makes senseles to put all the data simply to an ordinary joint phase curve. However, the curves at the "bright" state show a smaller scatter as compared to two curves in the "faint" state. These two curves are upper ones at the V-R diagram because of the "mean brightness -- color" dependence mentioned above.



\section{UBVRI Photometry}
\subsection{Color gradients}

The light curve obtained on January 17, 2004 at the AZT-11 telescope of CrAO is shown in the upper part of Fig.8. It shows two unequal minima, onto which are superimposed possible quasi-periodic oscillations and flickering. At the deeper minimum, the photon counting statistics is much worse, which causes an increase of observational errors.

The color indexes (middle part of the Fig.8) show that at minima, the emission becomes "bluer" in a sense of decrease of the color indexes. The amplitude of orbital variations increases with wavelength. The slopes are dU/dR=0.23(1), dB/dR=0.43(1), dV/dR=0.78(1), dI/dR=1.18(1).
This is opposite to the behavior of eclipsing cataclysmic variables with an accretion disk [55].
Qualitatively this situation may be explained by a "redder" spectral energy distribution of the accretion column than that of the white dwarf.

\subsection{Principal Component (PC) Analysis (A)}

Andronov et al. [12] have applied the PCA to determination of statistically significant number of sources of variability and their characteristics in 3 cataclysmic variables with different degree of influence of the magnetic field onto accretion. 
This method is an effective tool to determine components of variability, which have different spectral energy distributions, i.e. the wavelength dependencies of the amplitudes of variability. For two-color observations of the same accuracy and subtracted mean value, this method is known as the "orthogonal regression"; the principal components $U_{k\alpha}$ are
$U_{k1}=(x_{k1}+x_{k2})/\sqrt{2},$ $U_{k2}=(x_{k1}-x_{k2})/\sqrt{2},$ where $x_{k\alpha}$ are observations at point $k$ in channel (filter) $\alpha.$
So the first principal component shows common features in different channels (although with channel-dependent amplitude), and other PCs describe differences. The "color-brightness" diagrams, which are often used in astronomy (the Herzsprung-Russel diagram, diagrams for variable stars), are also a modification of the PCA with an aim to get more information on the objects. Polarimetry is also related to PCA, being applied to "ordinary" and "extraordinary" waves in the emission.

For the multi-color photometry, PCA is a supplementary statistically optimal tool to extract components of variability, which may be overseen if analyzing the colors only. Formerly, a method was proposed [54], which allowed to extract only the first component.

Applying the formalism of PCA [12] for the present data, we have found that the difference in the amplitude of the first principal component (U1) is much larger than of the others. It explains 84\% of overall variance in all colors. The second component U2 is responsible for 10\%, so the total  contribution of the rest three (U3, U4, U5) is only 5.8\%. Despite a scatter of the third component, it has obvious variations at the orbital time scale. The amplitude "signal/noise" ratio obtained using the "running parabola" scalegram analysis [4] is 26, 14, 8, 4 and 4 for U1--U5, thus we conclude that first 3 principal components are statistically significant. Relatively large values for U4 and U5 are due to larger errors at a second more deep minimum. 

The dependence of the matrix $V_{ij}$ on the number of channel (filter) $i$ and on the number of the PC $j$ is shown in Fig. 9 and is listed in the table below. The first (largest in the amplitude) PC shows both 3-hour and faster variability, as one may see in Fig.8. Its contribution monotonically decreases with the number of channel (and corresponding wavelength), as one may see from Fig.8. The ratios $V_{i1}/V_{41}$ (which are equal to ratios of the amplitudes of variability with a shape $U_{k1}$ are equal to 0.29, 0.47, 0.83, 1 and 1.20. The difference of these values from the ratios dU/dR, ..., dI/dR is present, because 3 components are statistically significant (for an exact linear dependence of the brightness variations in different colors, i.e. the presence of only one statistically significant component, the values in a pair for a given color must coincide).

The first PC $U_{k1}$ mostly resembles variations in I and R and much worse the variations in U, what is expected, looking on the increase of $V_{i1}$ with wavelength. Such a behaviour may be interpreted for this system, assuming that in BY Cam there are sources of emission with different spectral energy distribution: the dominating source of variability (first PCA) is due to cyclotron emission of the accretion column, so it has an amplitude, which strongly increases with wavelength. This resembles a classical polar AM Her with an accretion column ("infrared-dominated" variability), but quite opposite to the nova-like variable TT Ari ("ultraviolet-dominated" variability) with an accretion disk.

For the second PC in BY Cam, the dependence of $V_{i2}$ on $i$ has an opposite sign of slope than $V_{i1}$, i.e. $V_{i2}$ monotonically decreases with wavelength. This means that the second (in the amplitude) component has a "blue" spectral energy distribution, contrary to the "infrared" first PC. An examination of the curve (Fig. 8) shows that the second PC $U_{k2}$ is dominated by $\sim30-$min variability, whereas $U_{k1}$ is "spin-dominated".

\subsection{Scalegram Analysis}

The $"\Lambda-$scalegrams" [5] are shown in Fig. 10 for the original UBVRI data, for the corresponding color indexes and for the PCs.  All the curves have the same scale, but shifted to avoid an overlap. This method is some kind of the wavelet analysis, which is most effective for quasi-periodic oscillations, where one expects to determine a characteristic cycle length, rather than a constant period.
Contrary to the periodogram, wavelet and auto-correlation methods, which are higly affected by slow (periodic or aperiodic) trends, the $"\Lambda-$scalegram" is insensitive to polynomial trends up to degree 3. Thus, in a presence of strong asinusoidal spin variability (as in BY Cam and many other cataclysmic variables), we prefer to use this method to estimate effective (mean) amplitudes and periods (or cycles) of different components of variability.

 The height of the peak is proportional to the square of the "effective amplitude". The most pronounced peaks are seen for the scalegram for the first PC (U1) corresponding to "effective periods" of 0\fd122, 0\fd37, 0\fd21 (in order of decreasing height at the scalegram). The positions of the peaks are shown by vertical lines for comparison with that for other scalegrams. The highest peak corresponds to the most prominent "3-hour" variability. It is slightly less than a photometric period, what is expected for a non-sinusoidal shape of the curve [5]. The second period is close $P_{phot}/4,$ confirming a 4-peak structure of the light curve, which was pointed out by Pavlenko [40].

The third peak corresponds to $\sim 30-$min QPOs. Previously, QPOs in BY Cam were reported to occur at some phases with different estimates of their amplitude and time scale: 5 min [44], 10 and 15 min [50], 5-10 and 30 min [42, 51]. The fourth peak occurs at 0\fd0056=8 min and agrees with smaller periods mentioned above. However, its amplitude is smaller than for other peaks. Despite the peaks seem too small as compared to the main peak, the corresponding effective semi-amplitudes are large enough (0\fm58, 0\fm33, 0\fm24 and 0\fm11 for four peaks for U1, respectively). 

For U2, the peaks are of much smaller amplitude, but nearly at the same places, as for U1. The highest peak of $\Lambda(\Delta t)$ corresponds to 30-min component, similar to the scalegram for the filter U. This also is in an agreement with the conclusion of the "infrared" U1 and "blue" U2. The third U3 has a small amplitude, however, its variations are mainly concentrated to the "3-hour" peak.

In Fig. 10, the scalegram for 3-sec photometry (all 4 nights) obtained at 2.6m telescope ZTSh is also shown. Besides the dominating peak, it also has smaller peaks close to the same positions. The $"\Lambda-$scalegrams" for other runs show night-to night variability of the shape and corresponding amplitudes and "periods". This causes an overlap of the peaks and decrease of the ratio "peak-to-continuum". Thus an absence of well pronounced secondary peaks agrees with the former conclusion of significant variability of the characteristics of these brightness oscillations either during the run, or from run to run.

The height of the primary peak, which corresponds to the $3^{\rm h}$ variability, characterizes the amplitude of variations. Using the 4-harmonic fits, this value was determined for all runs, which sufficiently cover the photometric phase, and will be discussed in the last section in more detail. We have splitted the "all" data sets for the BOAO and K-380 telescopes into the "bright" and "faint" parts.

For the BOAO data, the height of the peak and a corresponding effective amplitude is larger for the filter R than for the filter V for "all", "bright" and "faint" data sets, in an agreement with the direct determination of the slope dV/dR$<1.$

The effective amplitude is by a factor of $\sim1.3$ larger at the "faint" state than at the "bright" one for both BOAO and K-380 data. The position of the peak in the "faint" state switches to the half of the period, indicating  
for the redistribution of the amplitudes of the harmonic components of the light curve.


\section{Characteristics of Individual Light Curves}

As one of the main results of the first "Noah project", Silber et al. [50] pointed out, that the phase of the sine wave fits for the $3^{\rm h}$ variability show an abrupt jump with a period of 7\fd26, which was interpreted as a half of the complete beat period [36]. So there are two switchings of the accreting column from one pole to another. If the 7\fd26 - period dominates, this means that two accreting poles at the white dwarf are separated by $\sim 180^\circ$ in longitude (with a similar latitude) and there will be two waves during one beat period. Diversity of the location or structure will lead to an unequality of these waves.

Pavlenko [40], based on 31 runs in the inactive state, has suggested a more complicated structure with 3 possible jumps at the phases $\varphi_{beat}=0.25,$ 0.55 and 0.78 with a possible interval of the "erratic" phase variations between the beat phases 0.0 and 0.25. The mean slopes of the O-C diagrams between the "main" switches at $\varphi_{beat}=0.25,$ 0.78 correspond to the "true" spin period $P_{spin}=0\fd138428.$ However, an additional jump  
at $\varphi_{beat}=0.55$ was interpreted as a switch between 3 (or even 4) accretion zones. In the beat phase intervals 0.25-0.55 and 0.55-0.78 for BY Cam, Pavlenko [40] had suggested a larger slope "phase/beat phase"  ${\rm d}\varphi/{\rm d}\varphi_{beat})$ (and thus an apparent photometric period, as $P=P_0(1+(P_0/P_{beat}){\rm d}\varphi/{\rm d}\varphi_{beat}),$ where $P$ is a trial value of photometric period, $P_0-$ its adopted initial value, $P_{beat} - $ beat period). 

It should be noted that in some nearly-synchronous polars the accreting region moves rather smoothly in longitude even while accreting onto a single pole, as the treading region drifts due the asynchronism. At first, such a model was discussed by Geckeler and Staubert [19] for another asynchronous polar V1432 Aql (see Andronov et al. [9] for results of the recent observational campaign on that object). Periodic variations of phases of the spin variability with the orbital phase were reported by Kim et al. [27] for another magnetic cataclysmic variable 1RXS J062518.2+733433 = MU Cam. The system is an intermediate polar, and such a dependence may argue for a diskless model of the system [32]. Despite the spin period in BY Cam is close to the orbital one, contrary to the intermediate polars, the asynchronous polars may be also called "nearly synchronous" intermediate polars. So asynchronous polars (or BY Cam-stars [33,23]) are "intermediate" between the classical and intermediate polars. The ratio of the spin period to the orbital one for the magnetic white dwarfs in cataclysmic binaries leads to a phenomenological classification [39], but it has a strong physical explanation [20,32,55]: for small (for a given accretion rate and orbital separation) magnetic moments, the thread radius is small, and there is a strong accretion disc without a strong evidence for accretion columns ("non-magnetic" systems, e.g. TT Ari). With increasing magnetic moment, increases the inner radius of the accretion disc, and accretion columns become stronger ("disc-fed" intermediate polars). Accretion disc may be disrupted in "disc-less" intermediate polars, but the stable spin period is shorter than the orbital one. Asynchronous polars evolve to a "phase-locked" synchronous rotation, which is typical for classical polars of AM Her - type. However, because of influence of other system parameters, sometimes systems with larger magnetic field of the white dwarf may stay at the "less-magnetic" branch of the classificatin described above.     

The main characteristics of the fits for BY Cam have been plotted as a function of the beat phase (Eq. 1) in Fig. 11. The statistically significant degree $s$ of the polynomial fit was determined for each run separately, and ranges from $s=1$ (sinusoidal variation) to $s=4.$
They and the light curve at the 10 equidistant phases are presented electronically in Table 9. Each point represent one night of the observations. As the majority of data were obtained in the photometric system R, we have used this sample for better statistics. Incomplete curves, where the data cover less than 0\fd1 have been excluded from the analysis, to avoid large errors of the fit in gaps. The resulting number of nights is 46.

To reduce the effect of inhomogeneous distribution of phases of observations, we have used the "orbit-averaged" mean $m_{mean}$ (i.e. the coefficient $C_1$ in the fit) instead of the "sample mean".

As one may see from the dependence of this "mean" value, the "bright" and "faint" states are well separated from each other. There are only 3 nights (from 45) inbetween these two groups. The transitions are much less frequent (i.e. 6\%) than the bright (n=11, 24\%) and "faint" (n=32; 70\%) states. The range of the mean values in the bright state $(\sim0\fm3)$ is much smaller than in the faint state $(\sim1\fm1).$ This is also valid separately for the maxima ("max" in Fig.11) and minima ("min"), arguing for a better stability of the light curve in the bright state.

Not unexpectedly, the maxima and minima in the bright state are brighter than these characteristics for the runs in the faint state. In the faint state, the "mean", "max" and "min" values show minima at the beat phases $\varphi_{beat}=0$ and 0.5. The initial epoch was chosen by Pavlenko [40] to correspond to the minimum of the sample mean brightness, so our result for the "orbital mean" just shows that the difference between the "sample" and "orbital" mean values is much smaller than the amplitude of the curve.

The "max" and "mean" values $m_{min}$ and $m_{max}$ are correlated $(\rho=0.91(6))$ with a sample mean values of 13\fm97(3) and 15\fm04(3). If (as for truly monoperiodic stars) there were no phase shifts, these values could be named as the "maximum" and "minimum" value of the mean phase curve. The slope d$m_{max}$/d$m_{min}$=0.74(5) is significantly smaller than unity, indicating an increase of the amplitude $(m_{min}-m_{max})$ on $m_{min}$. More usual form is the "amplitude $(m_{min}-m_{max})$ vs. $m_{mean}$". The correlation coefficient $\rho=0.40(14)$ is much smaller, but the correlation is still significant at the $"3\sigma"$ level. The corresponding mean values are 14\fm53(7) (13\fm59-15\fm46) and 1\fm07(3) (0\fm74-1\fm46) for $m_{mean}$ and $(m_{min}-m_{max}),$ respectively. The slope is positive d$(m_{min}-m_{max})$/d$m_{mean}=+0.17(6).$

To remove an obvious correlations between the brightness at some phases owed to variations of the mean brightness and to check the stability of the light curve, we have introduced two related parameters $"m_{max}-m_{mean}"$ and $"m_{min}-m_{mean}".$

For this further analysis, we have classified "intermediate" points as "bright", thus determining a dividing value of the mean brightness of R=14\fm21. 

The parameter $m_{max}-m_{mean}$ (from -0\fm6 to -0\fm3) for the bright state is much closer to zero than that for the faint state (from -1\fm0 to -0\fm4). This indicates a more extended hump for the bright state and more stable shape of the light curve. For the $(m_{min}-m_{mean})$ values, the range (0\fm4-0\fm7) for the "bright" state is also smaller than for the "faint" (0\fm29-0\fm88), but there is no significant systematic shift.

Both characteristics $(m_{max}-m_{mean})$ and $(m_{min}-m_{mean})$ have a large scatter. In the beat phase interval 0.03-0.15, there are 5 outstanding points (nights) with an amplitude, which is nearly twice larger than that at the nearby beat phases. The mean values for 46 runs are $\langle(m_{max}-m_{mean})\rangle=-0\fm57$ (from -0\fm83 to -0\fm31) and $\langle(m_{min}-m_{mean})\rangle=+0\fm52$ (from 0\fm29 to 0\fm88).
Unexpectedly, there is no correlation between these two characteristics. If one will define $\tilde{m}_{mean}=(m_{max}+m_{min})/2,$ then the values $(m_{max}-\tilde{m}_{mean})=-(m_{min}-\tilde{m}_{mean}),$ and there will be an exact negative correlation. The absence of correlation argues for a high variability of the shape of the light curve.

The dependence of the photometric phases $\varphi_{ph}$ on the beat phases $\varphi_{beat}$ is shown at the bottom part of Fig. 11. In the previous studies, only one phase of the maximum was taken into account (one-harmonic fit by Silber et al. [50] or an "asymptotic parabola" fit by Pavlenko [40]). Assuming a complicated structure of the light curve and applying a 4-harmonic fit, we have used the phases of both maximum $\varphi_{min}$ and minimum  $\varphi_{max}$ and the phase difference between them. The last value is called "asymmetry" and is an important parameter listed in the General Catalogue of Variable Stars [30, 47].

The corresponding dependencies have following differences. The parameter  $\varphi_{max}$ has jumps at $\varphi_{beat}=$ 0.78 and 0.23, as was found by Pavlenko [40]. However, all the parameters show outstanding points at corresponding diagrams, which are characterized by relatively small formal error estimates, and thus may be classified as statistically significant. The original brightness variations, which cause such pecularities, may be produced by cloud-type inhomogeneties of the accretion stream. Such "flare-like" events with an amplitude of hundreds millimagnitudes are distinctly seen at our curves. So a "filtration" of the "bad" curves producing outstanding points may be done, and more smooth curves could be seen the diagrams (Fig. 11). The criteria for such a "filtration" may be elaborated in future. In this work, we present results of modeling of all curves in extensive electronic tables as "observational facts", not removing any outstanding data.

In other words, present results show that the situation is much more complicated,
not because of bad observations, but because of the complexity of the system. 

At present, we may conclude that outstanding points really exist, but, if removing them, one may suggest good sequences for the remaining points. 
One can naively expect, that the light curve is dependent purely on the beat phase (thus on the periodically varying orientation of the magnetic field and the structure of the accretion flow). The scatter at the diagrams (Fig. 11) argues for additional valuable processes, particularly, mean magnitude changes, and large-amplitude flares.

For our data, we suggest that
the phases of minimum show the main jump at slightly larger phase $\varphi_{beat}=0.92$ for the "faint" data and possibly a more smooth variation for the "bright" data. There are obvious outstanding points, which correspond to the "global" lowest minimum, whereas at the phase light curve there may occur two local minima inside a "main" minimum. This does not challenge a general conclusion of switching accretion from one pole to another while "idling" of the white dwarf in respect to the secondary.

The mean value of $\varphi_{min}$ is 0.058(24), so the deviation from zero is formally not statistically significant. However, the initial epoch in Eq. (5) may be corrected to $T_{phot.min}(HJD)=2453022.276(3)$ (assuming the period of Pavlenko [40]). Our 46 minima timings are best fitted to the ephemeris
\begin{equation}
T_{phot.min.}(HJD) = 2 453 213.010(3) + 0.137123(3)\cdot E. 
\end{equation}
So the period value differs from that of Pavlenko [40] by $1\sigma,$ i.e. not significantly. The maximum phase shift due to the difference between the periods for our $446^{\rm d}$ of observations is only 0.01(1). This is comparable with the mean accuracy of the phase determination for a single curve. Thus a recomputation of all phases is not necessary.

The asymmetry $\varphi_{max}-\varphi_{min}$ has a drastic range of variations from 0.18 to 0.80, which is much larger than a mean accuracy estimate of 0.012. There is a slight evidence (at the $3\sigma$ limit) for a double-hump structure with minima close to the phases of jumps.

\section{Results}
\begin{itemize}
\item The main differences between the spin light curves in the "bright" and "faint" brightness states are: a) the mean color index V-R is larger for the "bright" state (this argues that the emission region associated with accretion has an "infrared" energy distribution); b) the "bright" light curve has a smaller amplitude; c) the "bright" light curve is more stable; d) the "bright" light curve has a smaller value of $|m_{max}-m_{mean}|$ indicating larger duration of the maximum (which is usually split into two humps).
\item the slopes dV/dR are more close to unity for the variability during a night, than for the mean brightness, indicating that the variability due to changing orientation is more "blue" than the emission from the mean accretion component and is more "red" than the total emission. This may be tentatively interpreted as a sequence with decreasing temperature "white dwarf$+$red dwarf", "accretion structure variable with orientation", and "accretion structure, which is not variable with orientation". However, at present, this conclusion is far from the theoretical modeling.
\item Contrary to the intermediate polars BG CMi and MU Cam, the variations in different colors show an increase of the amplitude with wavelength.
\item The second principal component of variability is characterized by a decrease of the amplitude with wavelength. The amplitudes for the 30-min, $P/4$ and $P$ variations are comparable, indicating that 30-min QPOs are more "blue" than the spin variability.
\item The characteristics of 46 pairs of spin maxima and minima for the season in 2003-2005 have been determined.
\item The optimal mathematical model for the four-hump light variations corresponds to a fourth-order trigonometrical polynomial. The dependences of these parameters on the phase of the beat period and on mean brightness are discussed. In an addition, the 30-min and 8-min quasi-periodic oscillations are present with a variable amplitude and effective cycle length.
\item The ephemeris for the orbital minima has been corrected.
\end{itemize}

The determined parameters may be used for the comparison to theoretic models of accretion in magnetic cataclysmic variables. The further multi-color monitoring is needed for studies of variability of the changes of the accretion structure and rotation of the white dwarf.

{\it Acknowledgements.
This work was supported by the Korea Astronomy and Space Science Institute Research Fund 2003 and 2007 and was partially supported by the Ministry of Education and Science of Ukraine. We thank anonymous referees for their fruitful comments.
E.P.P. thanks J.Babina for her assistance in observations.}

\newpage

\unitlength=1in
\begin{figure}
\begin{center}
\begin{picture}(3.24,3.39)
\put(0,3.39){\special{em:graph fig1.bmp}}
\end{picture}
\caption{Finding chart for BY Cam. The size of the field is 5.6', North is up and East is left. For a cross-identification, the stars from Henden and Honeycutt [22] (marked with "H"} and Sumner [52] (marked with "S") are shown.
\end{center}
\end{figure}

\begin{figure}
\psfig{file=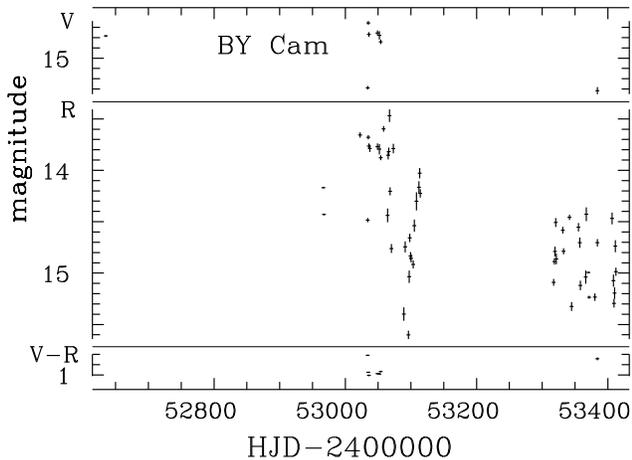}
\label{f2}
\caption{Dependence of nightly mean brightness in the filters V and R on time. R observations are taken for all 4 telescopes, whereas V are shown only for the 1.8m BOAO telescope. The error estimates of the correponding color index V-R is much smaller than of the original colors because of much smaller amplitude of its variability. The transitions between the "bright" and "faint" states usually take weeks, but sometimes even one day.}
\end{figure}

\begin{figure}
\psfig{file=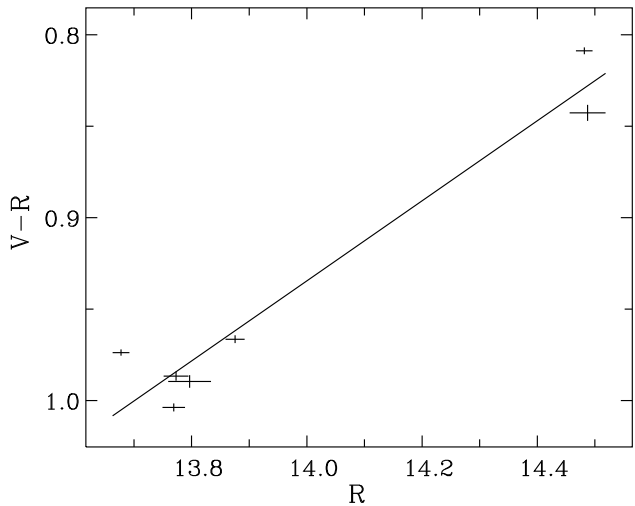}
\label{f3}
\caption{Dependence of the nightly mean color V-R on the brightness in R for the BOAO observations and the best fit (2).}
\end{figure}

\begin{figure}
\psfig{file=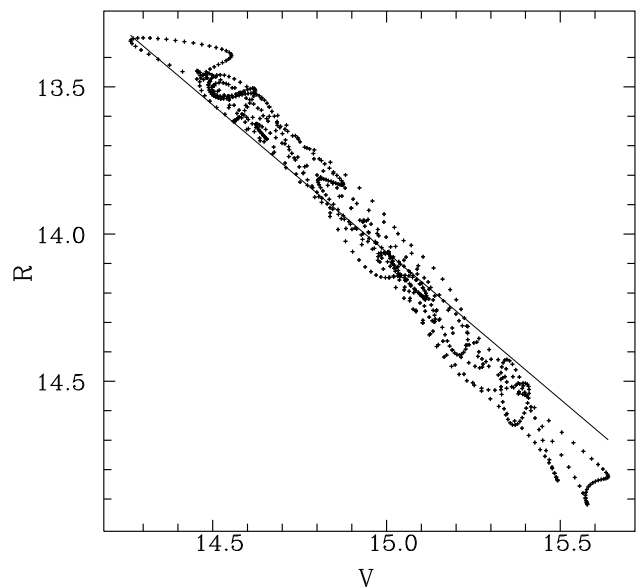}
\label{f4}
\caption{The "V-R" diagram for the 4-harmonic fits to the BOAO data. The phase distance between two subsequent points is 0.01P=2 min. The line $R-\bar{R}=V-\bar{V}$ with an unitary slope is shown for reference of individual slopes within 7 nights.}
\end{figure}

\begin{figure}
\psfig{file=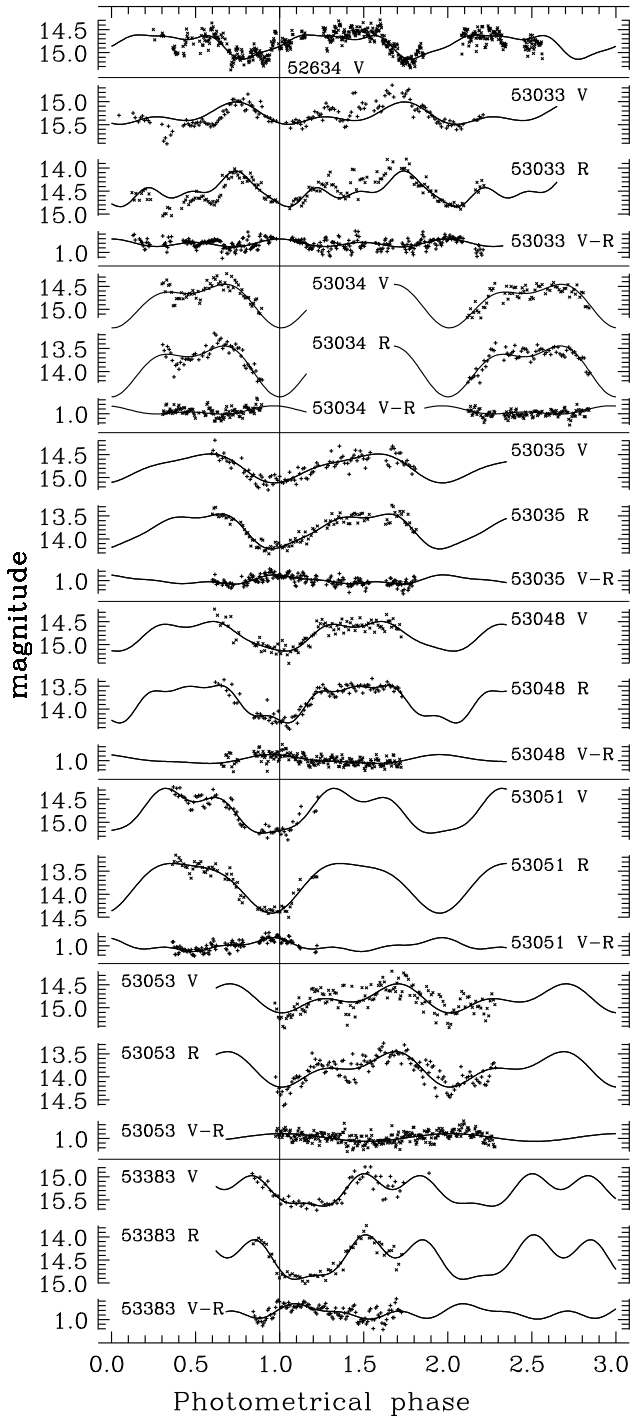}
\label{f5}
\caption{The individual light (V,R) and color index (V-R) curves (points) for the BOAO data. The original data are plotted once and are not repeated once per unit phase, but the fit is repeated for clarity.
The four-harmonic fits for individual runs are shown by solid lines. Strong night-to-night variability of the shape is present, with an easy visual separation of two curves observed at the "faint" state and five at the "bright state".
}
\end{figure}

\begin{figure}
\psfig{file=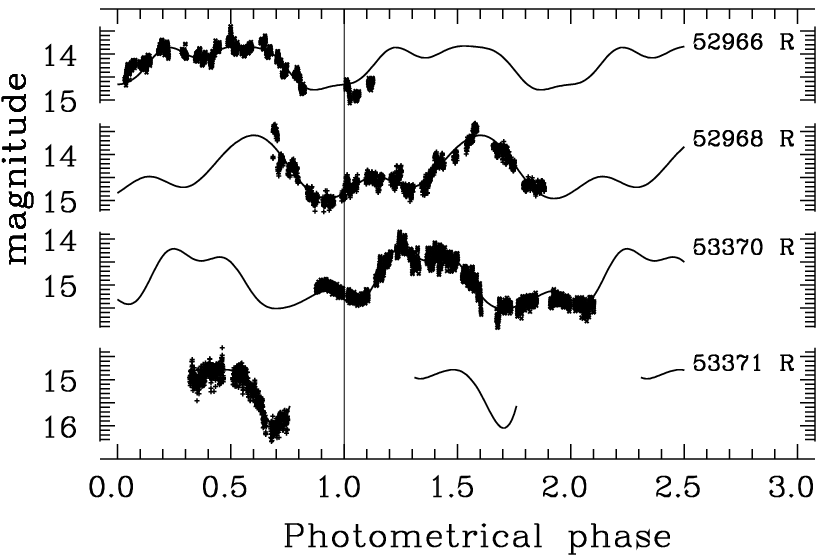}
\label{f6}
\caption{The individual light (R) curves (points) for the data obtained at 2.6m Shain telescope. As in Fig.5, The original data are plotted once and are not repeated once per unit phase, but the fit is repeated for clarity.
The four-harmonic fits for individual runs are shown by solid lines.}
\end{figure}

\begin{figure}
\psfig{file=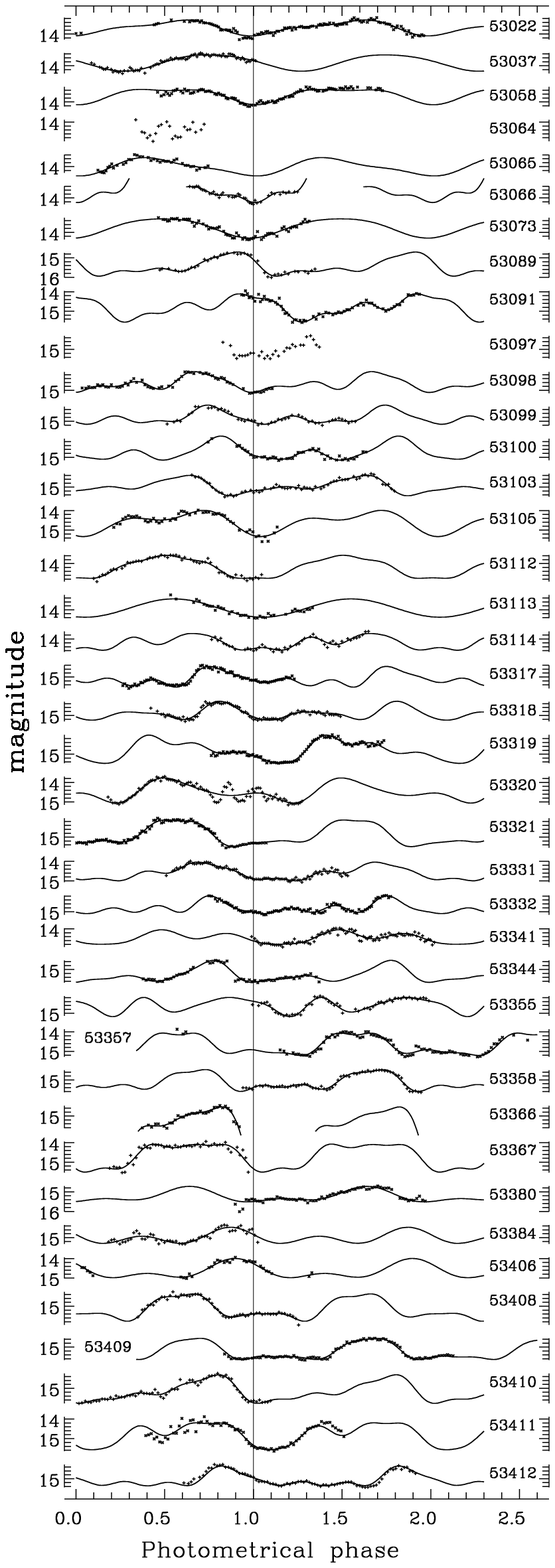,height=8.7in}
\label{f7}
\caption{The individual light (R) curves (points) for the data obtained at the telescope K-380. As in Fig.5, The original data are plotted once and are not repeated once per unit phase, but the fit is repeated for clarity.
The four-harmonic fits for individual runs are shown by solid lines.}
\end{figure}

\begin{figure}
\psfig{file=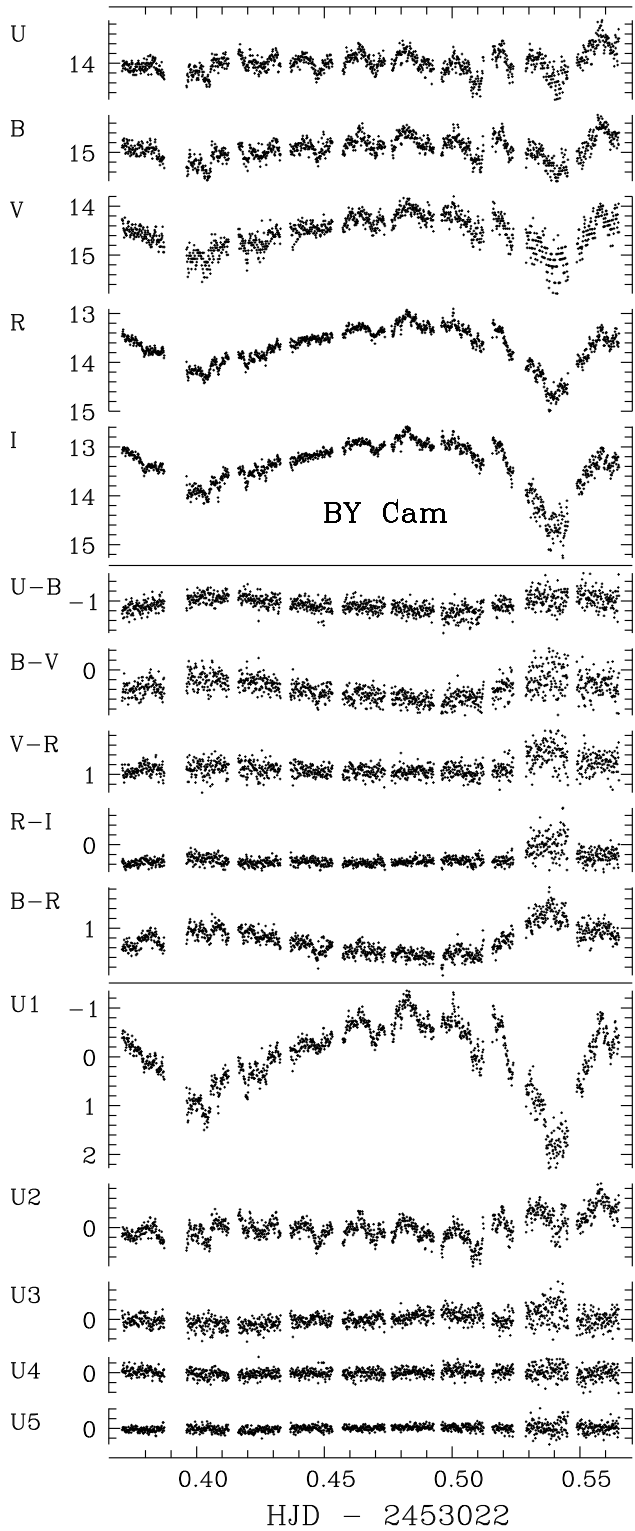}
\label{f9}
\caption{The individual light (UBVRI), color index (U-B, B-V, V-R, R-I, B-R) and principal component (U1,...,U5) curves (points) for the data obtained at the telescope AZT-11. The direction and scale are the same for all graphs. The interval between minor tics is 0\fm2.}
\end{figure}

\begin{figure}
\begin{picture}(3.43,1.83)\put(0,1.83){\special{em:graph fig9.bmp}}\end{picture}
\label{f10}
\caption{Dependence of the matrix of eigenvectors $V_{ij}$ on the number of channel $i$ and on the number of the principal component $j$ (see Andronov et al. [12] for a detailed description).}
\begin{tabular}{c|crrrr}
\hline
$j\backslash i$&1~~~~&  2~~~~  &  3~~~~ &  4~~~~ & 5~~~~\\
\hline
1 & 0.1541&   0.7440& -0.3797& -0.5132&  0.1233\\
2 & 0.2550&   0.4891& -0.0900&  0.7985& -0.2238\\
3 & 0.4455&   0.1977&  0.8123& -0.2333& -0.2196\\
4 & 0.5393&  -0.1452&  0.0022&  0.1459&  0.8166\\
5 & 0.6495&  -0.3835& -0.4335& -0.1528& -0.4687\\
\hline
\end{tabular}
\end{figure}

\begin{figure}
\psfig{file=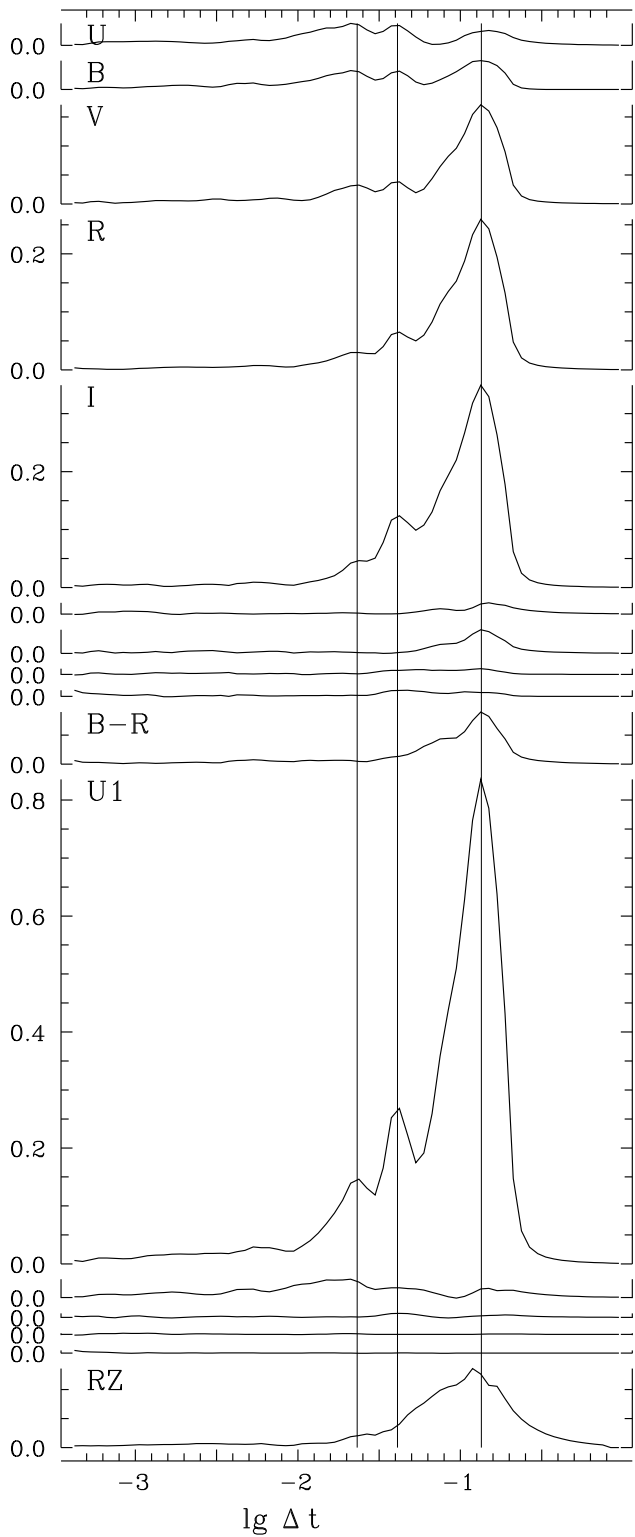}
\label{f11}
\caption{$"\Lambda"-$scalegrams [5] for the individual light (UBVRI), color index (U-B, B-V, V-R, R-I, B-R) and principal component (U1,...,U5) curves for the data obtained at the telescope AZT-11 and (bottom) for all 4 nights obtained at ZTSh. The vertical lines mark positions of 3 most prominent peaks at the scalegram for the first principal component U1.}
\end{figure}

\begin{figure}
\psfig{file=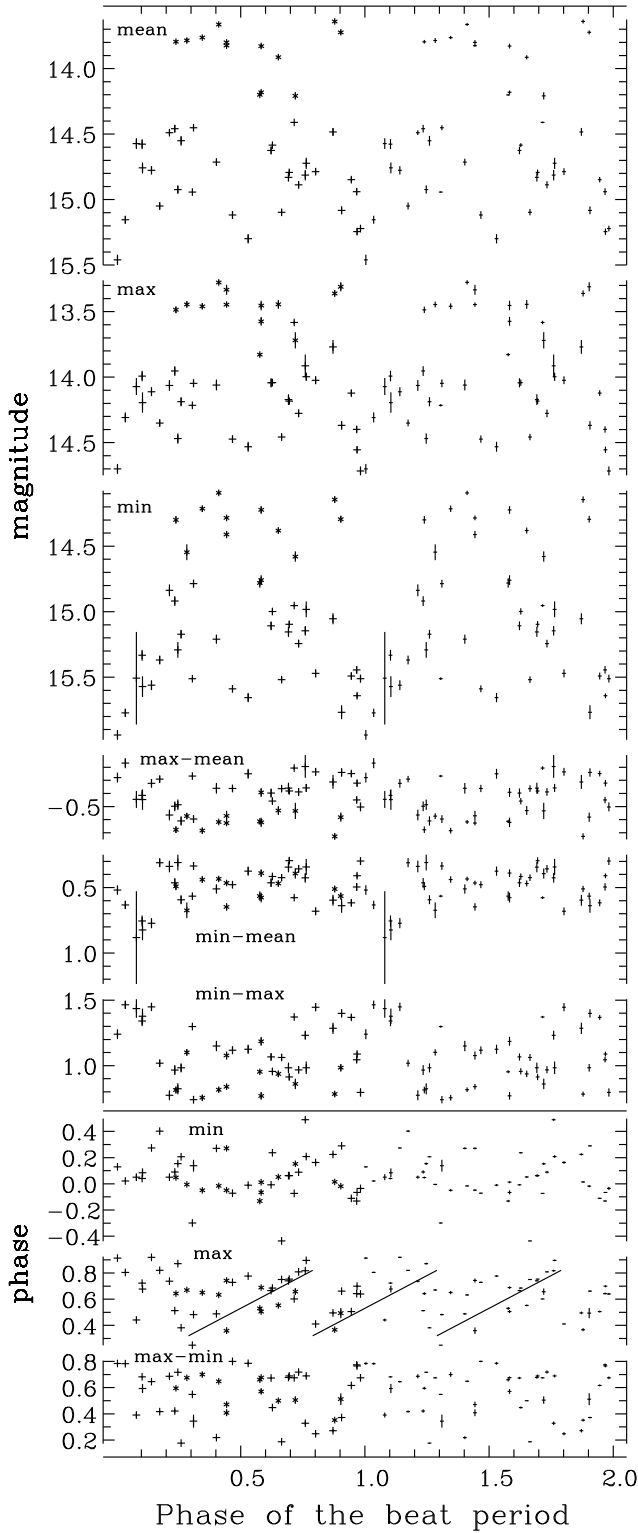}
\label{f12}
\caption{Dependence of the main parameters of the 4-harmonic fits to individual runs on the phase of the beat (synodic) period. Each point represent one night of observations.
For the phase range 0-1, the data are marked by asterics and crosses for the "bright" and "faint" states, respectively. The error bars are shown, but at the left part of the graphs, they are often smaller than the symbols, so at the right part (phases from 1 to 2) only error bars are shown. The slope of the inclined lines correspond to the spin period of the white dwarf, the jumps are interpreted as switchings of accretion from one pole to another.}
\end{figure}

\newpage

\begin{table}
\label{t1}
\caption{Brightness of BY Cam and the comparison stars from the BOAO observations (VR).
The values of $\sigma$ correspond to the r.m.s. scatter of a single observation in respect to the "artificial comparison star".
}
\begin{tabular}{cccccccccc}
\hline
Star&V&$\sigma_V$&R&$\sigma_R$\\
\hline
BY &14.867& 0.320& 13.992& 0.442\\
C1 &16.517& 0.004& 15.456& 0.006\\
C2 &12.982& 0.001& 11.815& 0.001\\
C3 &16.781& 0.015& 15.891& 0.027\\ 
C4 &13.375& 0.003& 11.490& 0.004\\
C5 &13.373& 0.004& 12.595& 0.004\\
C6 &14.793& 0.003& 13.690& 0.002\\ 
C7 &11.762& 0.001& 10.390& 0.001\\
C8 &12.106& 0.002& 11.027& 0.001\\
C9 &13.610& 0.007& 12.824& 0.007\\
C10&15.863& 0.010& 14.907& 0.017\\
C11&14.523& 0.003& 13.678& 0.002\\
C12&16.383& 0.017& 15.451& 0.041\\
C13&16.645& 0.010& 15.795& 0.011\\
C14&17.834& 0.030& 16.932& 0.056\\
C15&15.400& 0.007& 14.298& 0.005\\
C16&13.997& 0.005& 13.216& 0.003\\
C17&15.347& 0.010& 14.370& 0.004\\
C18&17.449& 0.034& 16.611& 0.058\\
C19&16.387& 0.007& 15.396& 0.009\\
C20&17.424& 0.016& 16.564& 0.030\\
C21&15.295& 0.007& 14.235& 0.005\\
\hline
\end{tabular}
\end{table}

\begin{table}
\label{t2}
\caption{Journal of observations of BY Cam:
JD- integer part of the starting Julian date (used for a legend of the run, e.g. 52634V);
HJD$_{start}$ - begin of observations (HJD-2400000);
telescope (Z=2.6m, K=1.8m, A=1.25m, P=0.38m);
filter;
exp - exposure in seconds;
N - number of observations;
 duration of the run in days.
In the electronic version of the table,
there are additional columns: HJD$_{finish}$ end of observations (HJD-2400000);
magnitude range for individual data points $m_{max}$, $m_{min}$;
nightly mean $\langle m\rangle$ and it's accuracy estimate $\sigma(\langle m\rangle);$
r.m.s. deviation of the single observation from the mean $\sigma(m);$
$\varphi_b$ - phase of the run center according to the "beat period" ephemeris
$T=2453089.2473+14.568E$ [40].
}
\begin{tabular}{ccccrrc}
\hline
JD   &  HJD$_{start}$ & Tel.& Filter& exp&   N & Duration\\
\hline
52634& 52634.93815& K   &   V   &  28&  523& 0.31725 \\
52966& 52966.46461& Z   &   R   &   3& 1536& 0.15014 \\
52967& 52967.47386& Z   &   R   &   3& 2089& 0.16433 \\
53022& 53022.26800& P   &   R   & 121&  119& 0.26880 \\
53022& 53022.37087& A   &   U   &  10& 1216& 0.19424 \\
53022& 53022.37087& A   &   B   &  10& 1216& 0.19424 \\
53022& 53022.37087& A   &   V   &  10& 1216& 0.19424 \\
53022& 53022.37087& A   &   R   &  10& 1216& 0.19424 \\
53022& 53022.37087& A   &   I   &  10& 1216& 0.19424 \\
53033& 53033.92933& K   &   V   & 105&  175& 0.29699 \\
53033& 53033.93985& K   &   R   & 105&  176& 0.28708 \\
53033& 53033.94088& K   &   V-R &  48&  325& 0.28544 \\
53034& 53034.92420& K   &   V   &  53&  126& 0.34822 \\
53034& 53034.92481& K   &   R   & 106&  125& 0.34820 \\
53034& 53034.92481& K   &   V-R &  48&  236& 0.34761 \\
53035& 53035.92456& K   &   V   & 105&  107& 0.16614 \\
53035& 53035.92516& K   &   R   & 105&  107& 0.16615 \\
53035& 53035.92516& K   &   V-R &  48&  200& 0.16554 \\
53037& 53037.36310& P   &   R   & 121&   80& 0.12700 \\
53048& 53048.95327& K   &   V   & 106&  100& 0.15222 \\
53048& 53048.95400& K   &   R   & 106&  100& 0.15205 \\
53048& 53048.95887& K   &   V-R &  48&  192& 0.14662 \\
53051& 53051.93498& K   &   V   & 105&   69& 0.11917 \\
53051& 53051.93560& K   &   R   & 106&   68& 0.11782 \\
53051& 53051.93560& K   &   V-R &  48&  126& 0.11782 \\
53053& 53053.93887& K   &   V   &  48&  142& 0.17939 \\
53053& 53053.93947& K   &   R   & 102&  140& 0.17935 \\
53053& 53053.93947& K   &   V-R &  46&  280& 0.17879 \\
53058& 53058.25650& P   &   R   & 121&  100& 0.17380 \\
53064& 53064.27290& P   &   R   & 181&   22& 0.05280 \\
53065& 53065.20350& P   &   R   & 121&   38& 0.08510 \\
53066& 53066.23430& P   &   R   & 121&   41& 0.08330 \\
53067& 53067.27850& P   &   R   & 156&    3& 0.00480 \\
53068& 53068.22690& P   &   R   & 158&    2& 0.01340 \\
53070& 53070.20850& P   &   R   &  43&   22& 0.02590 \\
53073& 53073.20320& P   &   R   & 130&   63& 0.11660 \\
53089& 53089.24730& P   &   R   & 156&   40& 0.12000 \\
\hline
\end{tabular}
\end{table}

\setcounter{table}{2}

\begin{table}
Table 2 (continued).\\
\begin{tabular}{ccccrrc}
\hline
JD   &  HJD$_{start}$ & Tel.& Filter& exp&   N & Duration\\
\hline
53091& 53091.23030& P   &   R   & 121&   66& 0.13730 \\
53096& 53096.29030& P   &   R   & 181&    8& 0.02170 \\
53097& 53097.24900& P   &   R   & 181&   30& 0.07460 \\
53098& 53098.23720& P   &   R   & 164&   57& 0.14690 \\
53099& 53099.26270& P   &   R   & 216&   49& 0.14550 \\
53100& 53100.25430& P   &   R   & 181&   43& 0.12200 \\
53103& 53103.25840& P   &   R   & 181&   64& 0.15660 \\
53105& 53105.25480& P   &   R   & 181&   50& 0.12630 \\
53108& 53108.30560& P   &   R   & 181&   12& 0.02850 \\
53112& 53112.23240& P   &   R   & 181&   55& 0.12970 \\
53113& 53113.25170& P   &   R   & 181&   37& 0.10970 \\
53114& 53114.24320& P   &   R   & 181&   48& 0.12120 \\
53317& 53317.38640& P   &   R   & 130&   87& 0.13250 \\
53318& 53318.36800& P   &   R   & 130&   91& 0.14640 \\
53319& 53319.37450& P   &   R   & 130&   88& 0.13340 \\
53320& 53320.39170& P   &   R   & 130&   98& 0.14990 \\
53321& 53321.46440& P   &   R   & 130&   89& 0.14640 \\
53331& 53331.40630& P   &   R   & 130&   90& 0.14020 \\
53332& 53332.39910& P   &   R   & 138&   86& 0.14040 \\
53341& 53341.34490& P   &   R   & 130&   93& 0.14130 \\
53344& 53344.55190& P   &   R   & 181&   57& 0.13600 \\
53355& 53355.19460& P   &   R   & 181&   58& 0.13640 \\
53357& 53357.19320& P   &   R   & 181&   81& 0.27090 \\
53358& 53358.20390& P   &   R   & 181&   63& 0.13770 \\
53366& 53366.21630& P   &   R   & 181&   33& 0.07490 \\
53367& 53367.15090& P   &   R   & 181&   49& 0.10710 \\
53370& 53370.39879& Z   &   R   &   3& 2612& 0.16804 \\
53371& 53371.41926& Z   &   R   &   3&  960& 0.05926 \\
53380& 53380.27430& P   &   R   & 181&   66& 0.14690 \\
53383& 53383.96821& K   &   V   & 167&   61& 0.12332 \\
53383& 53383.96911& K   &   R   & 168&  113& 0.29899 \\
53383& 53383.96911& K   &   V-R &  79&  117& 0.11962 \\
53406& 53406.20850& P   &   R   & 181&   32& 0.17780 \\
53408& 53408.30860& P   &   R   & 181&   58& 0.12420 \\
53409& 53409.19840& P   &   R   & 181&   82& 0.17660 \\
53410& 53410.18280& P   &   R   & 181&   68& 0.14810 \\
53411& 53411.19420& P   &   R   & 181&   72& 0.15320 \\
53412& 53412.18360& P   &   R   & 181&   82& 0.17680 \\
\hline
\end{tabular}
\end{table}

\begin{table}
\caption{Table of V observations (HJD-2400000, magnitude) obtained at the 1.8m telescope of the BOAO, Korea. {\em The table is very long, and is planed to be published as a text file electronically only}
}
\caption{Table of R observations (HJD-2400000, magnitude) obtained at the 1.8m telescope of the BOAO, Korea. {\em The table is very long, and is planed to be published as a text file electronically only}}
\caption{Table of R observations (HJD-2400000, magnitude) obtained at the 2.6m telescope ZTSh of the CrAO, Ukraine. {\em The table is very long, and is planed to be published as a text file electronically only}}
\caption{Table of brightness, color and principal components (HJD-2453022, U,B,V,R,I, U-B,B-V,V-R,R-I,B-R,U1,U2,U3,U4,U5) obtained at the 1.25m telescope AZT-11 of the CrAO, Ukraine. {\em The table is very long, and is planed to be published as a text file electronically only}}
\caption{Table of R observations (HJD-2400000, magnitude) obtained at the 0.38m telescope K-380 of the CrAO, Ukraine. {\em The table is very long, and is planed to be published as a text file electronically only}}
\end{table}

\begin{table}
\label{t8}
\caption{Color gradients dR/dV and dV/dR, number of pairs of interpolated VR measurements n, mean magnitude in V and R, correlation coefficient and the ratio to its statistical error (for individual nights, bright and faint state and data with a substructed corresponding nightly mean). {\em The table is planed to be published as a text file electronically only.}}
\end{table}

\begin{table}
\caption{The characteristics of the statistically optimal (up to 4-harmonic) fits to the individual runs and their error estimates:
$\varphi_{beat},$ $t_{mean},$ $m_{mean},$ number of the statistically significant degree of the trigonometric polynomial $s$, unbiased estimate of the "unit weight error" $s_0,$ r.m.s. accuracy of the fit $s_C$ $\varphi_{max},$ $m_{max},$ $m_{max}-m_{mean},$ $t_{max},$
$\varphi_{min},$ $m_{min},$ $m_{min}-m_{min},$ $t_{min},$ $\varphi_{max}-\varphi_{min},$ $m_{min}-m_{max};$
the light curve $m(\varphi)$ for $\varphi=$ 0.0, 0.1,... 0.9. {\em The table is long and wide, and is planed to be published as a text file electronically only.}
}
\end{table}

\end{document}